\documentclass[10pt, conference]{IEEEtran}

   \usepackage[pdftex]{graphicx}

\usepackage{array}
\newcolumntype{C}[1]{>{\centering\let\newline\\\arraybackslash\hspace{0pt}}m{#1}}
\usepackage{multirow}

\usepackage{amsmath}
\usepackage{todonotes}
\usepackage[ruled,linesnumbered]{algorithm2e}

\usepackage{listings}
\usepackage{color}
\usepackage{array}
\newcolumntype{L}[1]{>{\raggedright\let\newline\\\arraybackslash\hspace{0pt}}m{#1}}
\newcolumntype{C}[1]{>{\centering\let\newline\\\arraybackslash\hspace{0pt}}m{#1}}
\newcolumntype{R}[1]{>{\raggedleft\let\newline\\\arraybackslash\hspace{0pt}}m{#1}}
\usepackage{balance}

\graphicspath{{figs/}}

\begin{document}
\bstctlcite{IEEEexample:BSTcontrol}
\title{Leveraging Transprecision Computing for Machine Vision Applications at the Edge}

\author{\IEEEauthorblockN{Umar Ibrahim Minhas, Lev Mukhanov, Georgios Karakonstantis, Hans Vandierendonck and Roger Woods\\
School of Electronics, Electrical Engineering and Computer Engineering\\
Queens University Belfast, Belfast, UK\\}}



\maketitle

\begin{abstract}
Machine vision tasks present challenges for resource constrained edge devices, particularly as they execute multiple tasks with variable workloads. A robust approach that can dynamically adapt in runtime while maintaining the maximum quality of service (QoS) within resource constraints, is needed. The paper presents a lightweight approach that monitors the runtime workload constraint and leverages accuracy-throughput trade-off.  Optimisation techniques are included which find the configurations for each task for optimal accuracy, energy and memory and manages transparent switching between configurations. For an accuracy drop of 1\%, we show a $1.6\times$ higher achieved frame processing rate with further improvements possible at lower accuracy.

\end{abstract}

\vspace{-0.1cm}
\section{Introduction}
There is a growing need to execute complex machine vision applications on edge devices but they have their challenges. For example, a drone may simultaneously perform obstacle detection, people identification and crowd management where frame rate, accuracy requirements and active task state could need to vary at runtime. Edge devices need to operate under strict computational power and memory constraints. 

It is challenging to process the varying workloads on fixed hardware, while maintaining an algorithm's operating accuracy and a high Quality of Service (QoS) given as fraction of frames processed at an incoming rate. Machine vision algorithms can be tuned for the underlying hardware at reduced accuracy, but this does not reflect the variation of a system load at runtime \cite{fang2018nestdnn}, \cite{gao2018dynamic}. We can either conservatively design the algorithm to provide a low accuracy for a high system load or provide a high accuracy at low system load. Along with the varying system load, the dynamic parameters of a system, such as available memory or energy budget available for a specific task, may also vary.

Here, we propose a new design approach that provides the best trade-off between the QoS parameters of machine vision applications and usage of resources. This enables us to run efficiently such applications on the edge devices under strict operating constraints. We achieve this by developing a runtime system that dynamically selects the parameters for the selected Neural Networks (NN) models; these include bitwidth of floating-point instructions and the overall accuracy used by the applications to meet the required QoS, energy and memory restrictions. Our runtime system finds the best parameters for a specific NN from 3000 configurations identified in a previous study~\cite{scheidegger2019constrained}.

To find the optimal parameters for the required QoS under specific memory and energy constraints, we profile 3000 NN configurations on an edge device and then create various optimisation objectives, e.g. minimisation of execution time, power/energy consumption or maximisation of accuracy. We apply Integer Linear Programming (ILP) to find at runtime the optimal combination of configurations for a specific NN, meeting varying constraints, such as maximum energy/memory consumption. Our runtime monitors all the specified constraints and tracks the applications scheduled for execution. It automatically finds the optimal configuration for a NN used by the running application and transparently switches to this configuration to maximise overall QoS.

The paper is organised as follows. Section~\ref{background} gives the motivation and background. Section~\ref{method} formulates the problem and outlines the framework. Section~\ref{results} describes the experimental setup and analyses results before concluding in Section~\ref{conclusion}.

\section{Background and Motivation}
\label{background}

For NN-based machine vision, the optimisation of resources (execution time, energy, memory) has become key. The NN search can be optimised statically to find those models that can provide an optimal configuration for the underlying hardware, but this fails to react to the dynamically varying application needs in multi-task execution environments. Such needs may include changing workloads and accuracy requirements, as well as varying memory/power/energy availability per task in multi-application scenarios and battery operated devices. To cater for such scenarios, we apply transprecision optimizations that vary dynamically arithmetic precision and algorithmic accuracy.

Previous research studies trying to find the best trade-off between accuracy and latency for NNs applied an integrated approach where a single large network is designed as a superset of smaller subset networks \cite{fang2018nestdnn}\cite{kang2019dms}.

Such an approach improves computation efficiency by runtime truncating or suppressing computations in certain segments of NNs~\cite{gao2018dynamic} .
Other work \cite{rao2018runtime} has proposed a reward evaluation and decision-based runtime routing that allows choice of paths reducing the amount of computations without accuracy loss. Authors in \cite{kang2019dms} have designed adaptive pruning and reorganisation of compute intensive convolution filters based on their importance. A recent study \cite{fang2018nestdnn} implements a large multi-capacity model incorporating sub-models of various sizes that share the parameters. They design a cost function to find the best resource-accuracy trade-off and associated configuration for each of the concurrent tasks.

Although \cite{fang2018nestdnn} allows trading of resources via appropriate network design and optimisation of operating configuration at runtime, the use of an integrated large network has its constraints. Apart from lowering the number of operations, these approaches suffer from memory utilisation and high power consumption of larger networks. The adopted single objective during the training process, i.e. latency vs accuracy, does not allow optimisation of any other objectives at runtime, such as memory, energy consumption. For example,  a recent study \cite{yang2017designing} has shown that weight or computation reduction may not linearly translate into energy savings, and NNs should be optimised at the training stage to guarantee considerable energy savings. Finally, the runtime pruning of a single network can create irregularity and sparsity in the network leading to a low computing performance \cite{kang2019dms}.

We take a different approach that uses multiple independent smaller networks to achieve the best resource-accuracy trade-off. The networks have been designed to cover a wide design space and provide broadly varying computing characteristics, such as power, memory, energy consumption and latency. We profile and select the optimal configurations to suit varying runtime objectives, optimised via design of an ILP-based problem definition and solution. A runtime environment allows transparent switching of network configurations.

\textbf{NN Models:} Classical approaches use a uniform large space architecture search to find less computationally complex models. This is time consuming and leads towards larger models. Recent work~\cite{scheidegger2019constrained} proposes narrow searches in different spaces and their aggregation to construct a complete search. It uses a configuration file to describe rules for narrow searches drawing samples in a biased way to satisfy the constraints for a new model, i.e., the model size, the number of floating-point instructions (FLOP number). The original topology is ensured and includes analysis of half precision arithmetic usage against accuracy/memory utilisation.
In this work, 3000 reduced-accuracy models are distributed across a range of sizes for 5 well known NNs, such as Densenet, MobilenetV2, Googlenet, PNASnet and Resnext. This enables the authors to obtain a smooth pareto-optimal (PO) front for different accuracy-latency trade-off and achieve much higher object detection accuracy, more than 7\% for half of the NN models, 
against the same memory constraints as the original models. Nonetheless, the main goal for this study was to develop the NN models that can be used in devices with different performance characteristics while in our study we aim to investigate use of these models to vary performance and energy efficiency on the same device as per the runtime constraints.

 \begin{figure}[t]
\centering
\includegraphics[trim={1.5cm 4.0cm 3.2cm 1.5cm},clip, width=3.5in]{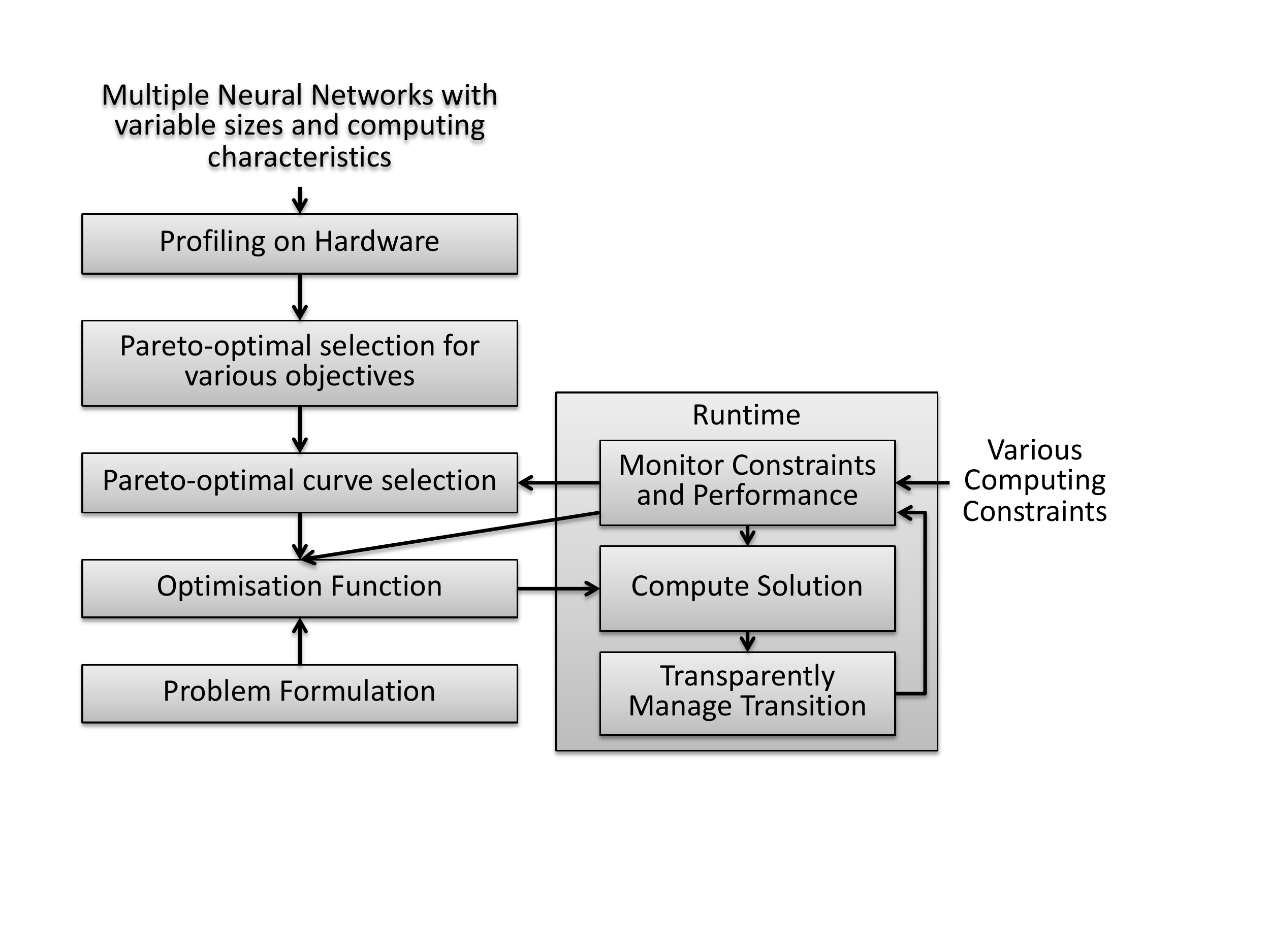}
\vspace{-.4cm}
\caption{Framework overview}
\label{fig_sys}
\vspace{-.7cm}
\end{figure}

\section{Proposed Framework}
\label{method}
The overall methodology presented at Fig.~\ref{fig_sys} makes use of a large design space generated via narrow searches using various segments of NN models, thereby introducing broad variability in generated models. It exploits the design space to create PO points targeting different optimisation objectives as variation in the NN segments can create varying data flows, as well as spatial and temporal computational densities and with varying computing resources, models may behave differently. For example, a low-memory model may achieve the lowest execution time on a memory constraint device, while a lower temporal computation complexity may provide the best energy efficiency for energy constraint environments. 

The first step thoroughly profiles all models for all of the different parameters that may need to be optimised or constrained during runtime. The second stage creates PO curves with different parameters, enabling a targeted design space that can be optimised further at runtime as per the varying objectives and constraints. We define the optimisation function using ILP, allowing runtime tunable constraints to be set as well as variation of optimisation objectives. We also design a runtime environment that enables transparent switching of various networks. It monitors runtime varying constraints, chooses appropriate PO selection, manages inputs to the ILP solver for the optimal solution and enables transparent switching of the networks while enabling the maximum QoS.

\subsection{Profiling of Neural Networks}

Using the generated models, execution time, power and memory utilisation is profiled both using the standard \textit{PyTorch} and Nvidia \textit{TensorRT} library models, a closed source library that optimises the execution for the underlying platform. For truncated networks, this is key as truncation can result in lower resource utilisation. The \textit{TensorRT} library first optimises the model and stores the resultant implementation, called \textit{engine}, which is an encoded file that needs to be read and decoded as an engine build process at runtime.

Although the underlying hardware provides support for half precision execution, the gains in execution time may not be significant if there is not enough parallelism, particularly for truncated lightweight networks. However, the lower resource utilisation may lead to lower power utilisation. Similarly, half precision requires lower storage space and thus lower overhead for loading and decoding engines at runtime.

\subsection{Pareto-optimal Selection}
PO selection identifies points which provide an optimum trade-off between two or more parameters, one of which is always accuracy while others may include execution time, power and memory. Whilst it reduces the runtime overhead, there is a trade-off in terms of the quality of resulting solution as it discards the design points which do not provide an optimal point based on the input parameters. To overcome this, different PO selections are created and used at runtime as per the optimisation objective. 

For a standard QoS metric, two parameters, accuracy and time, are sufficient but for hardware constrained devices, power and memory constraints may also apply. Thus, these configurations are introduced during PO selection. In one, we apply both time and the selected parameter during PO selection while in the other, we only consider the selected parameter and accuracy - offering a small search space and better optimisation of parameters against accuracy.

\subsection{Optimisation Function}
Description: Let us a consider a set of $N$ dynamically varying tasks $T = \{t_{1}, t_{2},... t_{N}\}$ to be executed on a device. Dynamic operation means that the workload per task can vary with time including going inactive. We define the workload as the required frames per second (FPS) which for the given task set, T, can be given by $F = \{f_{1}, f_{2},... f{n}\}$ where $0 \leq f_i \leq 30$. Furthermore, each task, $t_{i}$, has multiple versions where each version's implementation can provide varying throughput against accuracy, power consumption (energy) and memory utilisation. For simplicity and without loosing any advantages, we combine the implementations targeting algorithmic truncation and arithmetic precision into a single set which for task $t_{i}$ can be represented by $V_{i} = \{vi_{1}, vi_{2},... vi_{M}\}$. Different tasks can have different number of possible versions, however, we consider them to be the same here. Thus for each set $V_{i}$, we have an associated set of execution times, $T_{i}$, accuracies, $A_{i}$, power consumption, $P_{i}$, and memory usage, $M_{i}$.  

 In an ideal scenario, all tasks can run at the maximum accuracy while achieving the required FPS. However, for higher workload under low computing power, the aim is to select an appropriate version for each task that maximises the average operating accuracy, $A_{avg}$, for all tasks while maintaining the maximum FPS and adhering to constraints such as the minimum threshold accuracy, total power and memory consumption. Moreover, the runtime objectives can also be changed to minimise power/energy/memory consumption while ensuring a minimum threshold for parameters not being changed. 

The ILP problem formulation aims to find a set 
$V_{optimal} =\{vo_{1}, vo_{2},...vo_{N} \} \mid vo_{i} \in V_{i}$
i.e. one configuration each for all tasks, to be executed. If the corresponding execution time per frame, accuracies, power, energy and memory consumption of the found set of versions are represented by sets $T_{optimal}$, $A_{optimal}$, $P_{optimal}$, $E_{optimal}$ and $M_{optimal}$ and elements in each are represented by small case of the same letter and subscript $o$, the constraints for the problem are defined by:

\begin{itemize}

\item $\sum_{i=1}^{N} t_{o-i} \times FPS_{i} \leq 1 $ - all frames for all tasks should be processed in 1 second.

\item $ x > T_{th} \forall x \in T_{optimal} $ - tasks execution time should be higher than corresponding thresholds to control task priorities and resource allocation.

\item $ x_{i} > A_{th-i} \forall x \in A_{optimal} \mid 1 \leq i \leq N $ - tasks individual accuracies should be higher than corresponding thresholds.

\item $ x > P_{th} \forall x \in P_{optimal} $ - peak power consumption should be less than the threshold.

\item $\sum_{i=1}^{N} e_{o-i} =\sum_{i=1}^{N} p_{o-i} \times t_{o-i} \times FPS_{i} < E_{th} $ - total energy consumption in a second should be less than the threshold.

\item $\sum_{i=1}^{N} m_{o-i}  < M_{th} $ - total memory consumption of all tasks should be less than the threshold.
\end{itemize}
where the subscript $th$ defines the threshold set by the user for various parameters.

The joint memory consumption of all DNN models is used as they are simultaneous kept in memory. This avoids reloading the models on each context switch. The thresholds for different parameters are user defined, can be applied at any given time and are runtime variable. For example, the individual task's requirements for the execution time and accuracy can be defined for only a selection of tasks or completely omitted.

The optimisation objective can be dynamically selected from either of these 3:
\begin{itemize}
\item Maximise average accuracy, i.e., sum of accuracies for all tasks
\item Minimise total memory consumption
\item Minimise total energy consumption per second
\end{itemize}

Accuracy and energy provide direct performance improvement. Minimisation of memory can enhance performance for other co-executing tasks, not directly targeted by the optimisation function. Each objective can be achieved while keeping constraints on others. Constraints can also be set on the same parameters that are being optimised. For example, a minimum memory usage per task can be set or combined for all tasks, while the optimisation objective is to minimise memory. This is for scenarios where the designer knows that a reduction in memory usage can help overall performance to a certain limit.

\subsection{Runtime System}
The runtime monitors the changing constraints, generates new configurations and implements transparent switching of configurations. It is lightweight and responsive while acting to make sure optimal temporal utilisation of computing resources. 

In a stable system state, the optimal networks for all tasks are already loaded into the main memory and the frames for all tasks are being processed with the maximum QoS. The runtime monitors all of the system and task constraints. When there is a change in these, runtime goes into a transition state to switch to a new optimal configuration, while allowing the tasks to continue executing in the current configuration.

During the transition state, the runtime calls the ILP solver to find the next optimal combination. In the rare cases where a solution cannot be found, the runtime uses a user-provided priority to iteratively reduce the allowed FPS per task by 1. In this scenario, if there is a minimum time allocation constraint set for any task, it checks if that particular task is already within its limit and avoids further FPS reduction, by switching to a simpler but less optimal heuristic. Eventually, the ILP solver generates the optimal combination of configurations to be used for each task. The runtime then checks which tasks require a change in configuration. This gives an overhead related to the time spent on copying the model into main memory and the TensorRT desearlisation process which decodes stored models to make it available for use by the TensoRT.

To reduce this overhead, the runtime uses a buffer to load a new configuration in the background. Whenever a transition is started, the runtime allocates the time as needed, for the tasks which do not need a configuration change. The remaining time is equally divided across tasks which require configuration change and an FPS is selected which can be accommodated in that time slot, even if lower than required; a new configuration is loaded in parallel. As soon as it is loaded for a task, the runtime switches to it while updating the execution time and FPS. The remaining time is re-evaluated and allocated to the tasks for which configuration load is pending. The process continues until all of the tasks have their configurations updated, at which point the runtime exits the transition state. Typically, the transition state lasts less than a second.

To compare the performance and to allow usage where the ILP solver cannot find a solution, the runtime also implements other heuristics. These include: \textit{Fair FPS} which tries to allocate time per task that maintains same FPS for all tasks; \textit{Fair Time} which allocates the same time slot to all tasks and; \textit{Greedy} which allocates the requested time to the first and the remaining time to the second task and so on.

\section{Results}
\label{results}
\subsection{Experimental Setup}
In our experiments, we use NVIDIA Jetson Nano platform, which consists of Quad-core ARM A57 @ 1.43 GHz CPU and 128-core Maxwell GPU. It has 4 GB 64-bit LPDDR4 main memory and a class 10 ScanDisk micro-SD card with up to 170MB/s read and 90MB/s write speeds. The runtime was created in Python3.8, the NN architectures defined using PyTorch v1.7 and then optimised using Nvidia TensorRT v3.0.

For evaluation, we run in parallel up to 5 independent tasks (Tasks 1 - 5), where each task represents one of the original NN (described in Section \ref{background} A). We run each task while varying parameters, such as the workload FPS, accuracy requirements, etc. We limit the size of frames for each workload by 32x32 pixels. The measurements are provided at a granularity of 1 second.

Further analysis of benchmarks is provided via analysis of variation in parameters including execution time, power and memory using the \textit{TensorRT} library, as well as half precision number representation. Table~\ref{table_usecases} gives the average and maximum improvement over the baseline of the single precision standard computation; it shows that \textit{TensorRT} can speed up execution by up to $18.8\times$, at up to $7\times$ higher power usage as the resource usage efficiency increases. The memory consumption varied inconsistently, i.e, it was higher for some cases when using \textit{TensorRT} while lower in others. Furthermore, although the execution time for half precision is similar to single precision on average, it can provide better performance in some cases. Moreover, it provides more significant improvement over power usage (50\% or less on average for some of the networks) owing to lower resource utilisation as well as lower memory resulting in more than 50\% lower engine build times. The provided measurements highlight important information for designers looking to optimise lightweight neural networks on Jetson Nano platform.

\begin{table*}[]
\centering
\caption{Speedup for TensorRT and half precision}
 \vspace{-.2cm}
\label{table_usecases}
\begin{tabular}{|C{1.6cm} |  C{1cm} C{1cm}  C{0.8cm} C{0.8cm} | C{1cm} C{1cm} C{1cm} C{1cm} C{0.8cm} C{0.8cm} C{0.8cm} C{0.8cm} |}
\hline
\multirow{2}{*}{Network}&\multicolumn{4}{c}{Tensorrt} & \multicolumn{8}{c}{Half Precision} \\
\cline{2-13} 

 & Speedup Avg & Speedup Max & Power Avg & Power Max 
& Speedup Avg & Speedup Max & Memory Avg & Memory Min & Power Avg & Power Max  & Engine Avg & Engine Min\\
\hline
Densenet & 5.5 & 13.3  & 3.0 & 7.0
&1.1 & 1.9 & 0.8 &0.6& 0.8 & 0.5 & 0.4 & 0.1 \\

MobilenetV2 & 4.4 & 13.6  & 0.6 & 1.2
& 1.1 & 2.0 & 0.8 & 0.5 & 0.8 & 0.3 & 0.4 & 0.1 \\

Googlenet & 10.9 & 18.8  & 0.4 & 0.9
&1.1 & 2.2 & 0.8 & 0.5 & 0.5 & 0.3 & 0.4 & 0.1 \\

PNASnet & 2.6 & 5.6  & 1.1 & 3.7
&1.0 & 1.9 & 0.9 & 0.8 & 0.4 & 0.3 & 0.3 & 0.1 \\

Resnext & 7.4 & 12.8  & 0.4 & 1.3  
& 1.0 & 1.8 & 0.9 & 0.7 & 0.6 & 0.4 & 0.5 & 0.1 \\

\hline
\end{tabular}
\vspace{-.5cm}
\end{table*}

 \begin{figure}[t]
\centering
\includegraphics[trim={0.0cm 0.0cm 0.0cm 1.2cm},clip, width=3.2in]{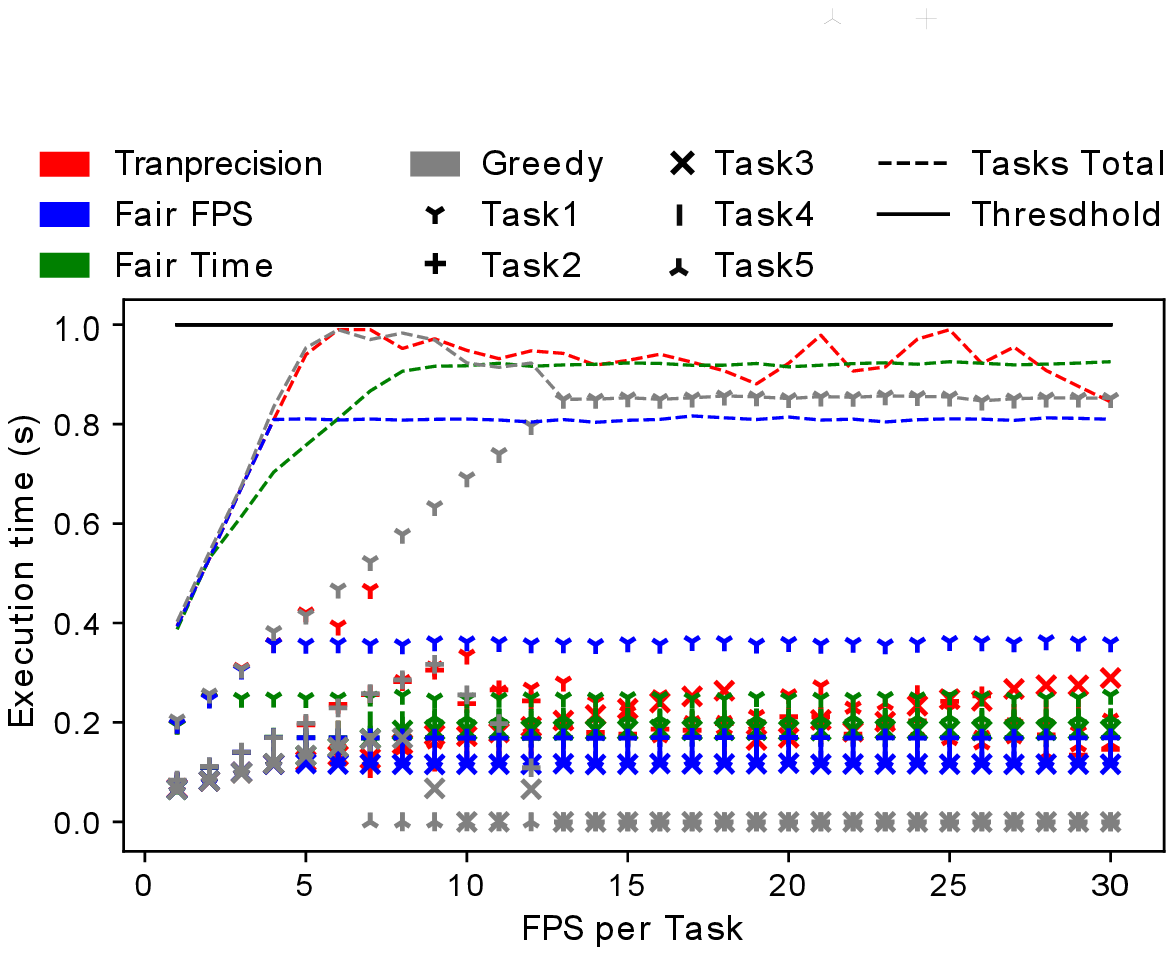}
\vspace{-.2cm}
\caption{Total and per task execution time for various heuristics}
\label{fig_time}
\vspace{-.4cm}
\end{figure}

 \begin{figure}[t]
\centering
\includegraphics[trim={0.0cm 0.0cm 0.0cm 0.0cm},clip, width=3.2in]{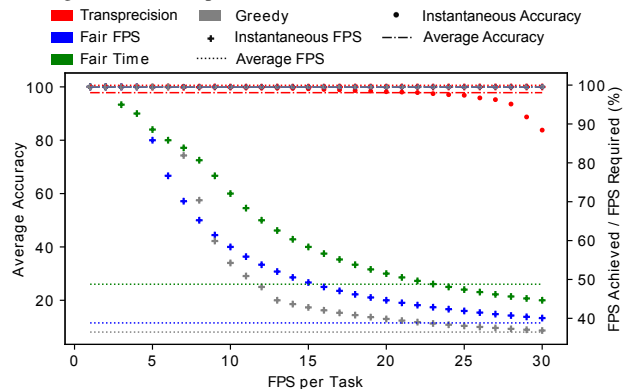}
\vspace{-.2cm}
\caption{Accuracy and achieved QoS for various heuristics}
\label{fig_qos}
\vspace{-.5cm}
\end{figure}

 \begin{figure*}[!htb]
\centering
\includegraphics[trim={0.0cm 0.0cm 0.0cm 0.0cm},clip, width=7.0in]{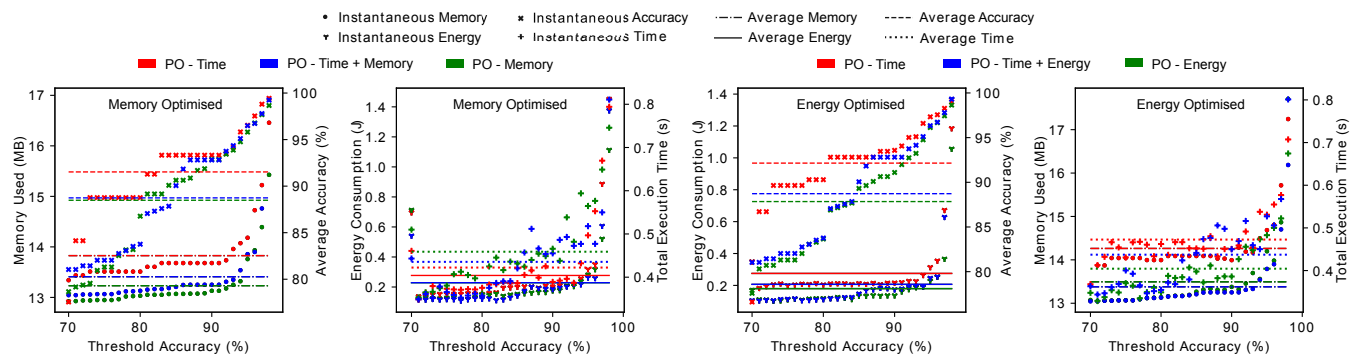}
\vspace{-.2cm}
\caption{Memory and energy optimisation against varying threshold accuracy}
\label{fig_acc}
\vspace{-.5cm}
\end{figure*}

\subsection{Changing FPS per Task}
We provide analysis for various heuristics use at runtime including the designed transprecision-based solution that looks to maximise accuracy while maintaining a 100 \% frame rate, the Fair Time distribution, the Fair FPS per task and the Greedy heuristic. For all approaches other than transprecision-based, the time to solution is negligible and does not affect the achieved FPS. 

For the transprecision-based approach, the ILP solver can take between 80 - 180 ms with an average engine build time of 20 - 80 ms for different NN and with a maximum of 400 ms. On average, the engine updates required per iteration are for 1 - 2 tasks. As mentioned earlier, the engine update is done in parallel and does not stall the processing. The only effect is that for the task requiring engine update, a lower number of frames may be processed. To simplify things, the time to find solution is also taken from the budget for the tasks whose engines need updated. The loss in FPS due to these overheads is considered when reporting the achieved FPS. However, the significance of both solution and engine update times depend on how often the solution needs to be recalculated and the engines are updated. For our experiments with a solution recalculation time of 5 s, i.e., incoming parameters changing every 5 s, the overheads were less than 0.06\%. 

Firstly, we look at the execution time per task and total execution time per second for various heuristics for varying FPS per task. Fig.~\ref{fig_time} shows that the non-optimised heuristics saturate the achievable FPS per task early within the available time budget and stay constant after that. For Greedy, some tasks do not get ever executed after a certain FPS, as the first task uses the whole time budget. All heuristics are able to finish processing in real-time before the deadline (1 s) showing that the pre-emptive scheduling is effective in this case, more so for transprecision execution which is able to fill the 1 s slot to a higher percentage due to its flexibility. We kept a 50 ms guard band per second, to account for CPU-GPU synchronisation and communication rather than the standard deviation in multiple iterations of tasks execution once a connection has been established. 

Fig.~\ref{fig_qos} shows the variation in average accuracy (relative accuracy against the maximum possible for that task) and percentage of achieved FPS against required FPS per task with variation in FPS per task. As shown, non-optimised heuristics maintain a maximum 100\% accuracy, but FPS drops sharply with increase in FPS per task; it goes as low as 8.67\% for 30 FPS per task. \textit{Fair Time} performs the best with an average of 48.78\% and a minimum of 20\% achieved FPS. The transprecision-based solution is able to maintain 99.94\% FPS (the solution provides 100\% FPS but the drop is due to the ILP solver and engine load overheads as explained above) up to 30 FPS. This comes at a slight drop of accuracy with an average of 97.83\% and a minimum of 83.79\% at 30 FPS. Even at 99\% accuracy (at 15 FPS per task), the approach provides $1.6\times$ higher FPS as compared to the next best.

\subsection{Changing Accuracy}
We explore optimisation of memory and energy against a varying minimum threshold accuracy for all tasks in Fig.~\ref{fig_acc}. To optimise an objective such as memory, the ILP solver can use any PO selection, i.e., time, time + memory and memory only (or time, time + energy, and energy only). For time only, the selection finds the optimal points that provide best accuracy against time; for memory/energy, it uses points that provide the minimum for any particular accuracy irrespective of time; finally, for time + memory/energy, it gives the optimal points against time and then discards those which do not give the best accuracy against memory/energy values. We give two sets (2 figures each) for each of memory and energy optimisation and for all the parameters (for both memory and energy optimised) including average accuracy as well as total memory, energy and time consumption. 

The results show that appropriate PO selections improves the efficiency by optimising the underlying objective as a trade-off against accuracy, while maintaining above the threshold. For example, the memory only PO selection for memory optimisation can provide on average 4.1\% lower memory utilisation at 3\% lower accuracy, while the energy only can provide 50.2\% lower energy solutions at 3.7\% lower accuracy. The time PO still gives the highest average accuracy. However, the non-optimised parameters such as time consumption and either of energy or memory for memory and power optimised execution respectively, provide varying trends, i.e., a memory/energy optimised solution could provide a lower energy/memory utilisation, but it is not guaranteed.

 \begin{figure}[!htb]
\centering
\includegraphics[trim={0.0cm 0.0cm 0.0cm 0.0cm},clip, width=3.5in]{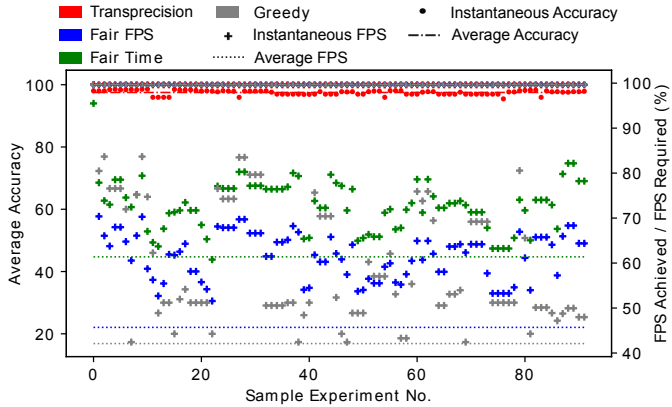}
\vspace{-.3cm}
\caption{Accuracy and achieved QoS for various hueristics  for run time varying constraints}
\label{fig_rand_qos}
\vspace{-.4cm}
\end{figure}

 \begin{figure}[!htb]
\centering
\includegraphics[trim={0.0cm 0.0cm 0.0cm 0.0cm},clip, width=3.5in]{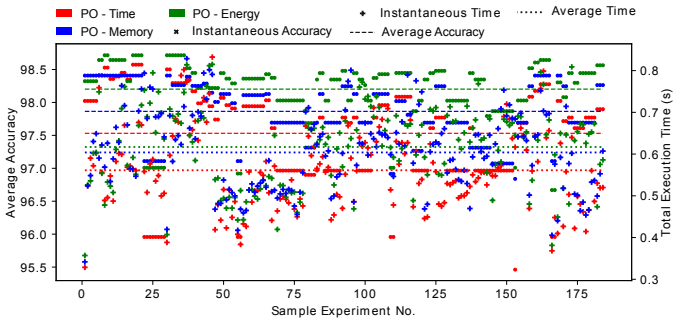}
\vspace{-.3cm}
\caption{Accuracy and execution time when executed with objective to maximise accuracy for random experiments}
\label{fig_rand_time}
\vspace{-.4cm}
\end{figure}

 \begin{figure}[!htb]
\centering
\includegraphics[trim={0.0cm 0.0cm 0.0cm 0.0cm},clip, width=3.5in]{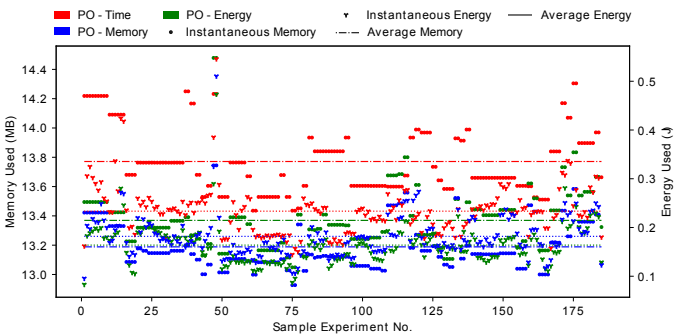}
\vspace{-.2cm}
\caption{Memory and energy objective optimisation executed with the appropriate pareto-optimal selections}
\label{fig_rand_mem}
\vspace{-.3cm}
\end{figure}

\subsection{Random Parameters}
Finally, we evaluate the heuristics under a more dynamic and constrained environment, by varying the individual threshold accuracies and FPS per task at runtime. We vary the total energy and memory available for all tasks, as well as the peak power. The parameters are varied randomly per iteration (every 5 s) while their range is selected to stress the system. For all non-optimised heuristics, all constraints are ignored to run the maximum accuracy models and only FPS is varied.

Firstly, we compare the achieved FPS and accuracy for various heuristics in Fig.~\ref{fig_rand_qos}. All heuristics other than the proposed solution run at maximum accuracy for all tasks; however, the FPS drops significantly with \textit{Fair Time} achieving maximum of 61\% while for \textit{Greedy}, it drops to 41\% on average. On the contrary, Transprecision is able to achieve 99.8\% FPS at slightly reduced 97.5\% accuracy. Next, we use different PO selections and different optimisation objectives on each. Fig.~\ref{fig_rand_time} shows contrasting results to Fig.~\ref{fig_acc} in that the optimised for time does not give the highest average accuracy. This is because in the current set of experiments, the energy and memory are more constrained than time and thus the PO selections based on energy/memory can offer better configurations to achieve highest accuracy. In this scenario, energy and memory optimal selections can provide 98.2\% and 97.8\% accuracy respectively as compared to 97.5\% for time-based selection. This suggests that in addition to optimising for a certain parameter, PO selections could be chosen, based on which resource is the most constrained. 

Finally, using the same experiments, instead of optimising for accuracy, we optimise for energy and memory while using appropriate PO selection for each and also comparing using only time-based selection. Fig.~\ref{fig_rand_mem} shows that it performs worse in terms of energy and memory consumption. On the contrary, the energy- and memory-based integrated optimisations can provide 49\% and 4\% lower energy and memory respectively, than the time-based selection.

\section{Conclusion}
\label{conclusion}
Whilst the operating accuracy of NN models for machine vision tasks on edge devices can be lowered, it is hard to find the optimal operating point by changing workloads, operating environment and optimisation objectives. We propose an approach for dynamic changing NN constraints and transparent switching NN configurations which demonstrates gains for various optimisation objectives while maintaining the maximum QoS.


\bibliographystyle{IEEEtran}
\balance
\bibliography{IEEEexample}

\end{document}